# Artificial Aesthetics: The Implicit Economics of Valuing AI-Generated Text

**Arbaaz Karim, Harvard College**

Aesthetic qualities command measurable premiums in traditional goods markets. However, it remains unclear whether users are willing to pay for such qualities in AI-generated text.

This paper estimates the willingness to pay for aesthetic attributes in large language model outputs using an online experiment with N = 117 participants. Participants evaluated responses from four anonymized models across academic, professional, and personal contexts, rated outputs along multiple dimensions, and submitted bids for access using a Becker-DeGroot-Marschak (BDM) mechanism.

We find no statistically significant relationship between perceived aesthetic quality and willingness to pay. While participants systematically distinguish between outputs and exhibit consistent preferences over stylistic features, these differences do not translate into higher monetary valuation.

Further analysis shows that aesthetic and functional attributes load onto a single latent factor, suggesting that users perceive “quality” as a unified construct rather than a separable aesthetic dimension. These results imply that, in current large language model (LLM) markets, aesthetic improvements function as baseline expectations rather than sources of price differentiation.

## Introduction

Large language models now reliably generate coherent, context-aware text consistent with human output, shifting attention from basic competence to aesthetic differentiation.[1-4] Reinforcement Learning from Human Feedback (RLHF) trains models to optimize for style, tone, and voice alongside grammatical correctness and factual coherence. In traditional goods markets, such aesthetic variations often command significant economic premiums. Yet, the monetary valuation of these attributes in the domain of synthetic media remains an unanswered question.

To address this gap, we depart from the productivity-focused methodologies of previous studies and turn to the foundational frameworks of economic valuation. The theoretical basis for valuing non-market attributes relies heavily on the work of Becker et al. (1964), who introduced incentive-compatible mechanisms for eliciting true willingness to pay.[5] Furthermore, we ground our analysis in the hedonic pricing framework of Rosen (1974), which posits that goods are valued for their utility-bearing attributes.[6] Under this framework, a complex good like a text output is viewed as a bundle of characteristics, and the market price is the sum of the marginal prices of these characteristics. Complementing this, Lancaster (1966) argued that consumers derive utility not from goods themselves, but from the specific characteristics they possess.[7]

In this paper, we study the willingness to pay (WTP) for AI-generated text using a BDM auction mechanism. We recruited 117 participants and tasked them with evaluating outputs from four leading large language models across three distinct contexts: Academic, Professional, and Personal use. Participants provided Likert-scale ratings of the outputs and subsequently placed monetary bids to purchase access to the LLM they selected for their own use. We also test whether users are willing to pay a premium for the "more beautiful" output, effectively measuring the shadow price of digital aesthetics.

We yield findings that challenge the assumption that “beauty” equates to higher value in the domain of LLMs.

The most notable result we report is a null result regarding the aesthetic premium. Our baseline regression estimates that a one-unit increase in perceived aesthetic quality yields a trivial and statistically insignificant increase in willingness to pay of $0.517. This result is robust across multiple specifications, including those that account for individual heterogeneity and zero-bidding behavior. Despite evidence that participants can and do distinguish between models based on aesthetic

criteria, these distinctions do not translate into monetary value.We argue that the lack of perceived economic value occurs because of the fundamental structure of beauty in AI text. We find that the construct of "beauty" is unidimensional. Principal Component Analysis reveals that a single latent factor explains 63.2% of the variance in user ratings. Functional attributes like "Clarity" and aesthetic attributes like "Prose Quality" load almost identically on this single factor, suggesting a halo effect where users collapse beauty into function. They do not view a text as "beautiful but useless" or "ugly but useful." Instead, they perceive a single dimension of general quality. Unidimensionality exposes a real-world limit to the theoretical applicability of the Lancaster (1966) model to this domain, as users effectively perceive only one characteristic rather than a bundle of separable traits.[7]

We document statistically significant heterogeneity in how aesthetic value is constructed, particularly through a phenomenon we term the "Definition Reversal." The specific drivers of choice shift dramatically based on the framing of the task. Participants who were primed to select a model for "beauty" surprisingly prioritized conciseness and efficiency, selecting shorter, more direct outputs. Conversely, participants evaluating models for "personal use" prioritized originality and prose quality. In the context of LLMs, the aesthetic ideal appears to be one of minimalist efficiency. Moreover, we find that the valuation of LLM style is highly uneven across the population. Low-WTP users show a significant sensitivity to aesthetics, while high-WTP users do not. We also observe a specific "brand premium" for ChatGPT, where users choosing this model exhibit a significant aesthetic elasticity not seen with other models.

## 1 Aesthetic Theory and Hedonic Pricing

The philosophy of aesthetics and the economics of attributes are two traditionally distinct domains that help construct a theoretical bridge for the intersection of LLMs and economic valuation. In this section, we provide philosophical theories that help explain user behavior and complement them with an empirical framework of digital pricing.

### 1.1 Philosophical Considerations

Just as Kaplan et al. (2020)[8] provided the scaling laws that govern the functional capability of these models (predicting the reduction of cross-entropy loss as a function of compute), we must ask what laws govern the aesthetic capability.

In Kant's *Critique of Judgment,* he characterizes the experience of beauty through the concept of "disinterested pleasure." He believes a true judgment of beauty must be devoid of any interest in the object's utility or existence.[9] One can judge a flower as beautiful as an appreciation of its form without desiring to eat it or own it. This creates a theoretical tension when applied to AI tools, which are inherently instrumental. If an LLM is a tool for productivity, can its output ever be judged with the disinterestedness required for true aesthetic appreciation? We approximate this distinction by comparing a "Beauty" frame to a "Personal Use" frame. If Kant's disinterestedness has an analogue in digital tools, we might expect different valuation patterns across these frames. In theory, the "Beauty" frame should elicit judgments based on form and style, divorced from the immediate utility of the text, whereas the "Personal Use" frame should trigger instrumental valuation.

Hume's landmark *Standard of Taste* grapples with the subjectivity of aesthetic judgment. He acknowledges that "beauty is no quality in things themselves," yet also argues that a consensus of "true judges" could establish a standard.[10] In the context of LLMs, RLHF is one institutional mechanism for aggregating the judgments of many "true judges." Models are fine-tuned on the aggregated preferences of thousands of human raters (presumably an expert in the domain), effectively industrializing Hume's standard. Statistically significant heterogeneity in how users rate and value the same text outputs would confirm subjectivity or context-dependency.

However, the question of whether there is a divergence between the pursuit of "veritas" in quality optimization by the LLM firms and the idiosyncratic taste of the consumer remains. We can expect a consensus among participants if the RLHF process has successfully encoded a universal standard of taste. Whereas, if taste remains subjective or context-dependent, we would expect to see statistically significant heterogeneity in how users rate and value the same text outputs.

### 1.2 The Economics of Digital Attributes

Rosen (1974) establishes the first set of hedonic pricing framework, which helps translate Kant and Hume's philosophical concepts into economic terms.[6] According to Rosen, goods are valued for their utility-bearing attributes. Take a car, for example. It is a bundle of horsepower, fuel efficiency, leather seats, and sunroofs. When you pay the market price of the car, it is the sum of the implicit prices of these attributes. Consumers inspect the bundle and pay a premium for the specific attributes that maximize their utility. This framework has been the foundation of quality-adjusted pricing for decades, allowing economists like Berndt & Griliches (1993), for example, to separate the value of a faster processor from the value of a larger screen in the pricing of computers.[11]

Aesthetic attributes often carry a high marginal cost and thus a significant price in the physical world. The car with a hand-stitched leather interior costs more to produce than one with cloth, and this scarcity supports its price premium. However, the digital economy introduces an anomaly where there is zero marginal cost. Digital assets can be reproduced at negligible cost. In the context of Generative AI, the marginal computational cost of generating a "beautiful" sentence is identical to that of generating a "utilitarian" one.

When style has effectively zero marginal production cost, it is an open question whether markets assign it a positive implicit price. Thus, an argument can be made for an "Instrumental Hypothesis" which would suggest that the user's relationship to AI is entirely instrumental. From this perspective, one only cares about the functional output of the tool; "beauty" being an epiphenomenon or perhaps even a contradiction in terms. If this is the case, one should find that the implicit price of the "aesthetic" attribute $A$ in the bundle is not significantly different from zero. In other words, one is maximizing a utility function that is only a function of the efficiency of information transfer. The other argument would be an "Aesthetic Hypothesis" that, as functional quality is commoditized, “beauty” becomes the primary axis of aesthetic differentiation. As the theoretical supply of serviceable text nears infinite, the relative scarcity of distinctive (or pleasing) text should drive its value proportionally higher. This study will provide a measurement for adjudicating between these two hypotheses in the economic sense of measuring the willingness to pay for $A$ (holding $F$ constant).

## 2 The Market for AI Text

In our test, we construct a controlled market for LLM output. We designed the setting to mimic the current real-world deployment of these tools while allowing for precise measurement of user preferences and valuations.

### 2.1 The Models (The "Firm")

The "supply side" of our market consists of four distinct Large Language Models. We selected leading models that represent the current frontier of generative capability: Gemini 3.0 Pro (Google), Grok 4.1 (xAI), Claude 4.5 Opus (Anthropic), and ChatGPT 5.1 (OpenAI). While the underlying architecture for all these models is based on the Transformer[12], they differ significantly in their fine-tuning methodologies and resulting "personalities." These models function as the "firms" in our market, each producing a differentiated product (text) in response to the same input (prompt).

Brand effects in the AI market are potent; thus, we anonymized the models (labeled A, B, C, D) during the evaluation phase to prevent brand bias from confounding the aesthetic evaluation. By blinding the models, we ensure that the WTP bids reflect the value of the content itself, not the trademark. Additionally, memory-enabling settings were turned off in the production of all outputs to ensure unskewed results from each respective model. We reveal the brand names in our post-hoc analysis to examine brand-specific effects and uncover whether there is intrinsic aesthetic value from the brand equity.

### 2.2 The Tasks

The "demand side" of our market is defined by the tasks users perform. Recognizing that the utility of text is highly context-dependent, we selected the Academic, Professional, and Personal contexts to cover a wide spectrum of modern knowledge work.

In the academic task, participants were asked to evaluate outputs that explained the concept of opportunity cost as if they were beginners. The specific prompt used was “Explain the concept of opportunity cost in economics as if I'm a beginner. Use examples to make it clear.”

In the professional task, participants evaluated email drafts addressed to their manager asking for career development advice. The specific prompt used was “Draft an email to my manager requesting a meeting to discuss career development. Sound professional but not overly formal.”

For the personal task, participants evaluated creative, informal writing that asked for help creating a travel itinerary. The specific prompt used was “I'm deciding whether to take a 10-day trip to Japan and China in January or to Australia and New Zealand. Help me figure out what would make the most sense.”

By varying the task, we ensure that our measure of aesthetic value is not limited to a single domain. This design mirrors the real-world marketing of these tools, which positions them as versatile assistants for work, study, and creative endeavors. The prompts used in evaluating these three domains were intentionally simple and undetailed to elicit variance in the outputs.

## 2.3 The Mechanism (The BDM Auction)

While survey methods (e.g., Likert scales) effectively measure attitudes, they are notoriously poor at predicting behavior due to the absence of cost. A user might say they value beauty, but are they willing to pay for it? The gap between stated preference and revealed preference is a well-documented phenomenon in economics. Hence, we rely on the use of a Becker-DeGroot-Marschak (BDM) auction to solve this issue.[5]

The Becker-DeGroot-Marschak (BDM) mechanism is frequently used to determine an individual's willingness-to-pay. It requires an individual to submit a bid for an item. A price is then randomly selected. If the submitted bid exceeds this price, the individual acquires the item and pays the randomly drawn price. If the bid is lower than the price, the individual does not receive the item and pays nothing.

This incentive makes the incentive of truth-telling a dominant strategy, unaffected by the individual's risk attitudes or whether they maximize expected utility. Published studies from Davis and Holt (1993), Kahneman et al. (1990), Rutstrõm (1998), and Shogren et al. (2001) use the BDM auction in incentive-compatible manners for non-random goods.[12-15]

In our experiment, after rating the model outputs, participants were given a $15 bonus allocation they could use to "purchase" premium access to their preferred model. The mechanism works as follows:

**Review:** Participants review the outputs from the four models.

**Bid:** Participants state the maximum amount ($0.00 to $15.00) they are willing to pay to secure the use of their chosen text.

**Random Price:** The system draws a random price from a uniform distribution between $0.00 and $15.00.

**Outcome:** If the participant's bid ≥ the random price, the transaction executes at the random price. If the bid < the random price, no transaction occurs.

Therefore, the dominant strategy for a risk-neutral participant is to bid exactly their true value. Bidding higher than one's true value exposes the participant to the risk of paying a price higher than their utility. Bidding lower risks losing a transaction that would have yielded a surplus. Importantly, we framed this purchase in a way that mimicked a micro-transaction or a subscription premium, a realistic scenario for how these tools are monetized.

While participants were informed of the mechanism ex ante, payments and allocations were not executed contingent on realized bids. As a result, the design violates the standard requirement of consequentiality. This implies that elicited bids should be interpreted as hypothetical or weakly incentivized valuations rather than fully incentive-compatible WTP. To the extent that incentive salience is reduced, estimated effects may be attenuated toward zero, particularly in the presence of low-stakes or preference-uncertain decisions. Accordingly, the null aesthetic premium should be interpreted with caution as a lower-bound estimate of valuation effects.

## 3 Data and Empirical Strategy

We collected data from 117 participants, resulting in 1,392 individual ratings across the four models and three contexts.

Participants were recruited on Prolific, a leading data collection marketplace, and restricted to respondents who (i) reported living in the United States, (ii) were at least 18 years old, and (iii) indicated that they had used an AI system before. This screening targets users of AI tools, though the sample is not nationally representative.

We introduce three layers of randomization into the experiment design. Participants were randomly assigned to either a personal-use frame or a beauty frame when prompted to select a model. Because assignment to the treatment group is random, the frame indicator is orthogonal to demographics, prior AI experience, and baseline preferences, which explains our interpretations of any differences in behavior across conditions as causal effects of the framing rather than of compositional differences.

To prevent question order bias within each participant, the order of the three contexts (academic, professional, personal) was randomized. Additionally, within each context, the order of the four model outputs was independently randomized. Every participant therefore rated all four models on all three tasks, but the sequence in which they encountered contexts and model outputs varied across subjects. This design reduces primacy and recency effects, learning effects, and simple fatigue patterns that might otherwise correlate with a specific model or context.

The models themselves were anonymized and presented as “Model A,” “Model B,” “Model C,” and “Model D.” The mapping between provider and label was fixed within a participant across all three tasks, so that “Model A” referred to the same underlying LLM each time that participant saw it. However, participants never saw the underlying provider names and were not told how many providers were involved. Because all participants in a given condition see the same pre-generated outputs for each model–context pair, and because model identity is masked and order is randomized, variation in ratings, choices, and willingness to pay can be interpreted as valuation of the content and perceived qualities of the text, rather than of brand reputation or a particular position in the sequence.

### 3.1 Data Construction

To quantify the attributes of the text, we constructed two primary indices based on the five dimensions rated by participants.

For each model–context pair, participants rated clarity, tone appropriateness, prose quality, conciseness, and originality on 1–7 scales, along with an overall aesthetic slider. Before constructing indices, each dimension was standardized within the sample. Principal components analysis is used later as a robustness check, but all main regressions rely on these straightforward averaged indices rather than factor scores.

**Functional Index (*F*):** This index is the simple average of standardized clarity, tone, and conciseness scores. These dimensions correspond to the utilitarian aspects of the text. A high score on this index implies that the text is easy to understand, appropriate for the context, and efficient in its delivery.

**Aesthetic Index (*A*):** This index is the simple average of standardized prose quality and originality scores. These dimensions correspond to the hedonic aspects. A high score here implies that the text is elegant, stylish, novel, or creatively distinct.

### 3.2 Summary Statistics and Raw Distributions

Figure 1 (Summary Statistics) reveals a distinct valuation hierarchy where functional competence significantly outweighs aesthetic novelty, evidenced by Clarity achieving the highest attribute score ($M = 6.17$) while Originality recorded the lowest ($M = 5.31$), confirming that users evaluate these tools primarily as engines for coherent communication rather than novel ideation. The utilitarian preference is reinforced by the Functional Index ($M = 5.96$) consistently outperforming the Aesthetic Index ($M = 5.54$), yet this high perceived competence fails to stabilize economic value, as demonstrated by the extreme volatility in WTP, where the standard deviation (SD= \$4.21) is nearly two-thirds the magnitude of the mean bid ($M$ = \$6.70).

| Variable | Mean | SD | Min |
|---|---|---|---|
| BDM bid ($) | 6.70 | 4.21 | 0.00 |
| Clarity (1–7) | 6.17 | 0.91 | 3.33 |
| Prose quality (1–7) | 5.77 | 0.98 | 3.00 |
| Tone (1–7) | 5.92 | 1.01 | 2.33 |
| Conciseness (1–7) | 5.78 | 0.99 | 2.33 |
| Originality (1–7) | 5.31 | 1.06 | 2.67 |
| Overall aesthetic (1–7) | 5.77 | 0.97 | 2.67 |
| Functional index (F) | 5.96 | 0.87 | 2.89 |
| Aesthetic index (A) | 5.54 | 0.90 | 3.00 |

Figure 1: Descriptive Statistics of User Ratings and Valuations

While the mean bid was \$6.70, a significant minority of participants (12.8%) bid \$0.00. This zero inflation suggests that for a segment of the population, the base utility of the text does not cross the threshold of monetization, or they perceive the marginal value of the "best" model over a default alternative to be zero. The disconnect between the normal distribution of ratings (perception) and the skewed distribution of bids (valuation) foreshadows our main result: users see the difference, but they don't necessarily pay for it. The median bid is significantly lower than the mean, indicating that a few high-value bidders are pulling up the average, while the "median voter" places a relatively low monetary value on the text.

### 3.3 Empirical Strategy

The regression specification we employ to estimate the aesthetic premium uses the following hedonic regression framework:

$$WTP_{ic} = \alpha + \beta_1 A_{ic} + \beta_2 F_{ic} + \gamma X_i + \varepsilon_{ic}$$

Where:

- $WTP_{ic}$ is the willingness to pay of participant $i$ for the chosen model in context $c$.
- $A_{ic}$ is the Aesthetic Index of the chosen model.
- $F_{ic}$ is the Functional Index of the chosen model.
- $\varepsilon_{ic}$ is a vector of controls including demographics (age, gender) and usage frequency.

The coefficient of interest is $\beta_1$. A positive and significant $\beta_1$ would indicate the existence of an aesthetic premium (that users pay more for text they perceive as more beautiful, holding functional quality constant). This baseline specification was chosen for its interpretability and direct mapping to Rosen's framework. We also employ a Tobit model to account for the left-censoring of the WTP data at zero (since users cannot bid negative amounts). Furthermore, we use quantile regressions to explore how the valuation of aesthetics varies across the distribution of willingness to pay, allowing us to see if "high rollers" value beauty differently than budget-conscious users.

## 4 Main Results

Our analysis of the WTP data finds that the market, under these conditions, does not reward a “premium” for aesthetic differentiation of model output.

### 4.1 The Aesthetic Null

Figure 2 estimates the marginal WTP for an extra point on the aesthetic index, holding functional quality constant. Across all specifications, the coefficient on the Aesthetic Index fails to achieve statistical significance.

| Variable | (1) Baseline | (2) +Frame FE | (3) +Interaction | (4) Full model |
|---|---|---|---|---|
| Aesthetic index ($A$) | 0.517 | 0.446 | 1.085 | 0.861 |
| SE | (0.650) | (0.660) | (0.838) | (1.067) |
| Functional index ($F$) | −0.329 | −0.303 | −0.404 | −0.123 |
| SE | (0.674) | (0.677) | (0.680) | (1.067) |
| Beauty condition | — | −0.547 | 5.565 | 6.450 |
| SE | — | (0.798) | (5.028) | (5.674) |
| $A$ × beauty frame | — | — | −1.102 | −0.753 |
| SE | — | — | (0.895) | (1.361) |
| R² | 0.006 | 0.010 | 0.023 | 0.024 |
| Observations | 117 | 117 | 117 | 117 |

Figure 2: Regression Estimates of the Aesthetic Premium

In Column 1 (Baseline), we regress WTP on the Aesthetic Index ($A$) and Functional Index ($F$). The coefficient for $A$ is 0.517 with a standard error of 0.650 ($p = 0.428$). This implies that a one-point increase in perceived beauty (on a 7-point scale) is associated with a 52-cent increase in willingness to pay, but this effect is statistically indistinguishable from zero. The functional index, $F$, also fails to reach significance in the joint model, likely due to the high correlation between the two indices, but even when $A$ is modeled alone, its predictive power on WTP is negligible ($R^2 < 0.01$).

The interaction with the experimental frame (Column 3) further deepens the puzzle. We would intuitively expect that participants in the "Beauty" condition (who were explicitly instructed to select the most "beautiful" model) would show a higher valuation for aesthetics. However, the interaction term ($A$ x Beauty) is negative (-1.102) and non-significant. This suggests that priming users to think about beauty does not increase their willingness to pay for it; if anything, it may dampen the valuation, a phenomenon we explore in Section 5.

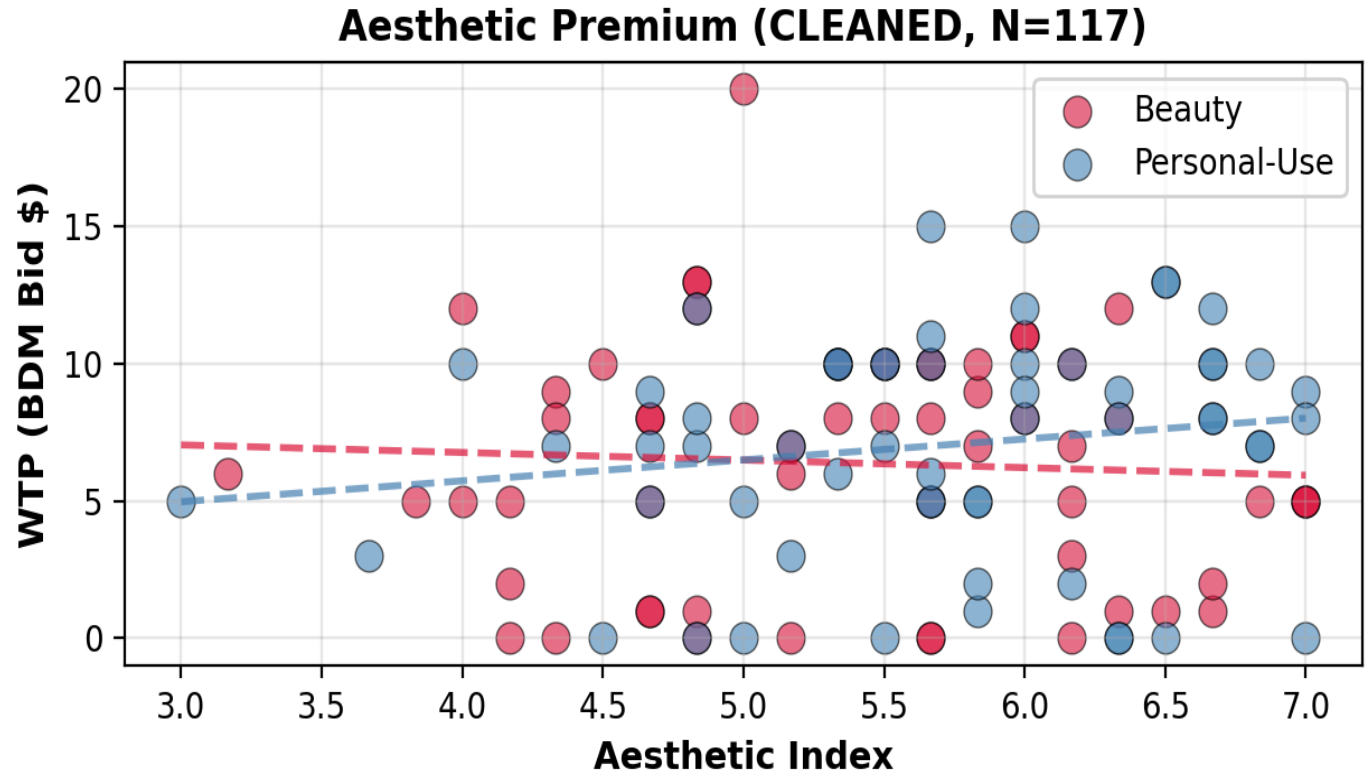

Figure 3: The Relationship Between Aesthetic Quality and Willingness-to-Pay

In Figure 3, we plot the relationship between the Aesthetic Index and WTP. The regression line is nearly flat, visually confirming the null result found in our models. The scatterplot's cloud of points with no discernible upward trend reinforces that for the average user, "better" prose does not equal "more valuable" text.

## 4.2 Residualized Aesthetics

A potential fault of the baseline model is the high correlation between the Aesthetic and Functional indices ($r = 0.743$), which could induce multicollinearity. To address this, we constructed a measure of "Pure Beauty" by regressing the Aesthetic Index on the Functional Index and taking the residuals. These residuals represent the variance in beauty that is orthogonal to function (the purely decorative or stylistic elements of the text).

| Predictor variable | β | SE | t-statistic | p-value |
|---|---|---|---|---|
| Pure beauty (residualized $A$) | 0.517 | 0.650 | 0.796 | 0.428 |
| Functional index ($F$) | 0.068 | 0.454 | 0.151 | 0.881 |
| Constant (intercept) | 6.294 | 2.731 | 2.304 | 0.023 |

Figure 4: Residualized Regression Summary Statistics

As shown in Figure 4, the coefficient on this residualized metric remains exactly 0.517 ($p = 0.428$), identical to the baseline specification. The stability in residualized and baseline coefficients confirms that the market price for pure aesthetics in AI output is indistinguishable from zero. Unlike the productivity gains observed in prior literature, where functional augmentation drives value, our results indicate that users treat "pure style" as a non-monetizable attribute. In the absence of functional improvements, consumers are unwilling to pay a premium for beauty, suggesting that the aesthetic component of generative AI is valued only insofar as it signals or correlates with functional competence.

## 4.3 Heterogeneity by WTP and Brand

Interesting patterns emerge when we examine the distribution of users. In Figure 5, we consider how estimated effects differ by user WTP decile. We find a significant effect at the 25th percentile of the WTP distribution. For these "low-WTP" users (those who are generally reluctant to pay), the coefficient on aesthetics is 2.263 ($p = 0.025$). This implies that for budget-conscious consumers, aesthetic quality is a significant differentiator. This group likely represents the "marginal consumer" who needs a reason to open their wallet at all. For them, basic functionality is expected (and thus priced at zero), but a truly beautiful output can trigger a positive bid. The "high-WTP" users (the 75th and 90th deciles) show no aesthetic sensitivity. Their willingness to pay is likely driven by functional necessity or deep-pocketed indifference to marginal quality differences.

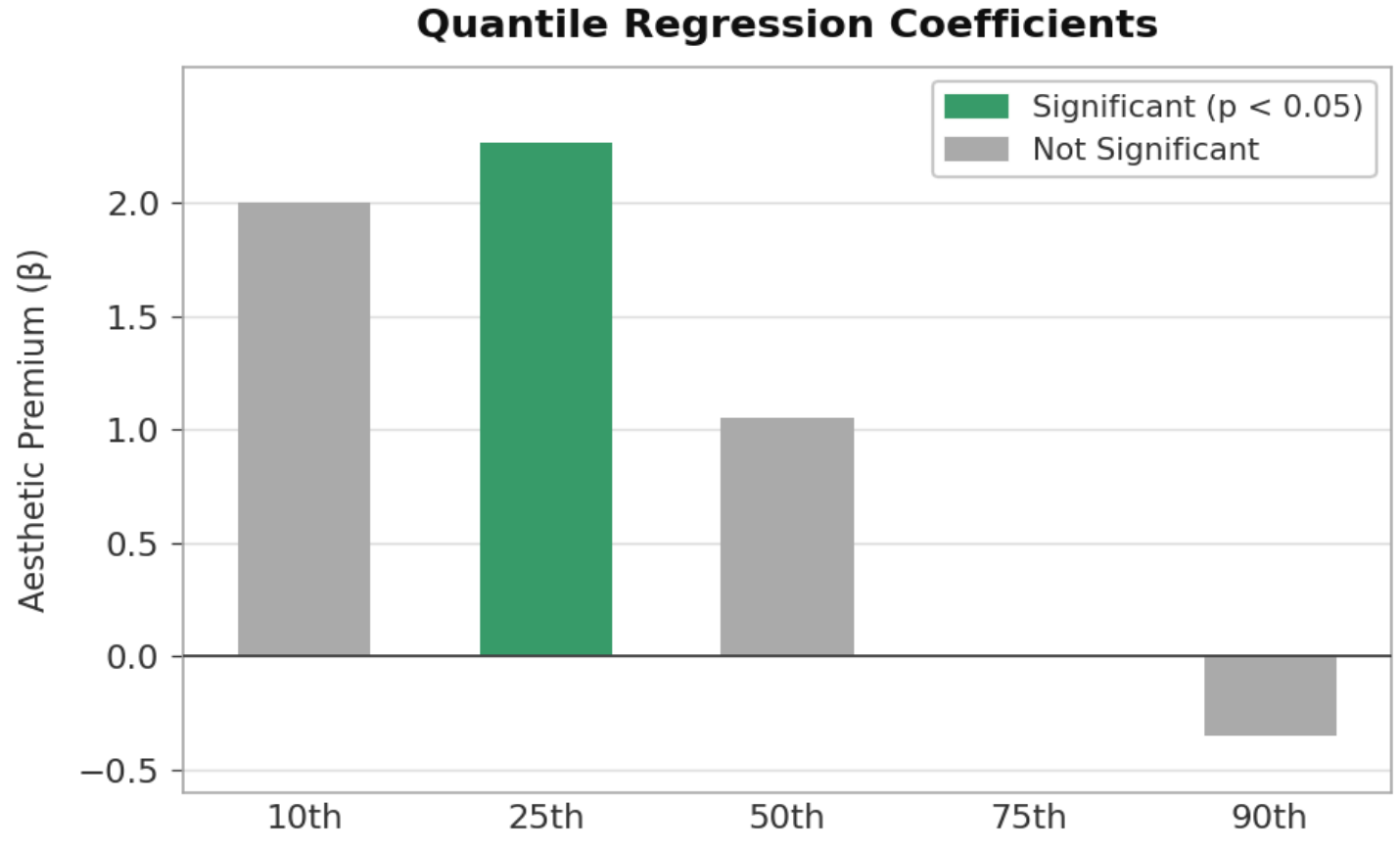

Figure 5: Quantile Regression Coefficients

Figure 6 highlights a striking anomaly with ChatGPT when we segment the sample to participants who chose specific models. When participants choose ChatGPT, the aesthetic premium becomes large and significant ($\beta = 3.774$, $p = 0.024$). This is a model-specific heterogeneity. Although our sample is too small to estimate these effects with high precision in all subsamples, the model-specific heterogeneity suggests that for certain brands and user segments, aesthetic style may matter more. This finding mirrors the "superstar" effects described by Rosen (1981), where small differences in talent (or in this case, style) can command large premiums in specific markets. It implies that ChatGPT has established a "brand aesthetic" that users recognize and value, distinct from the generic output of other models.

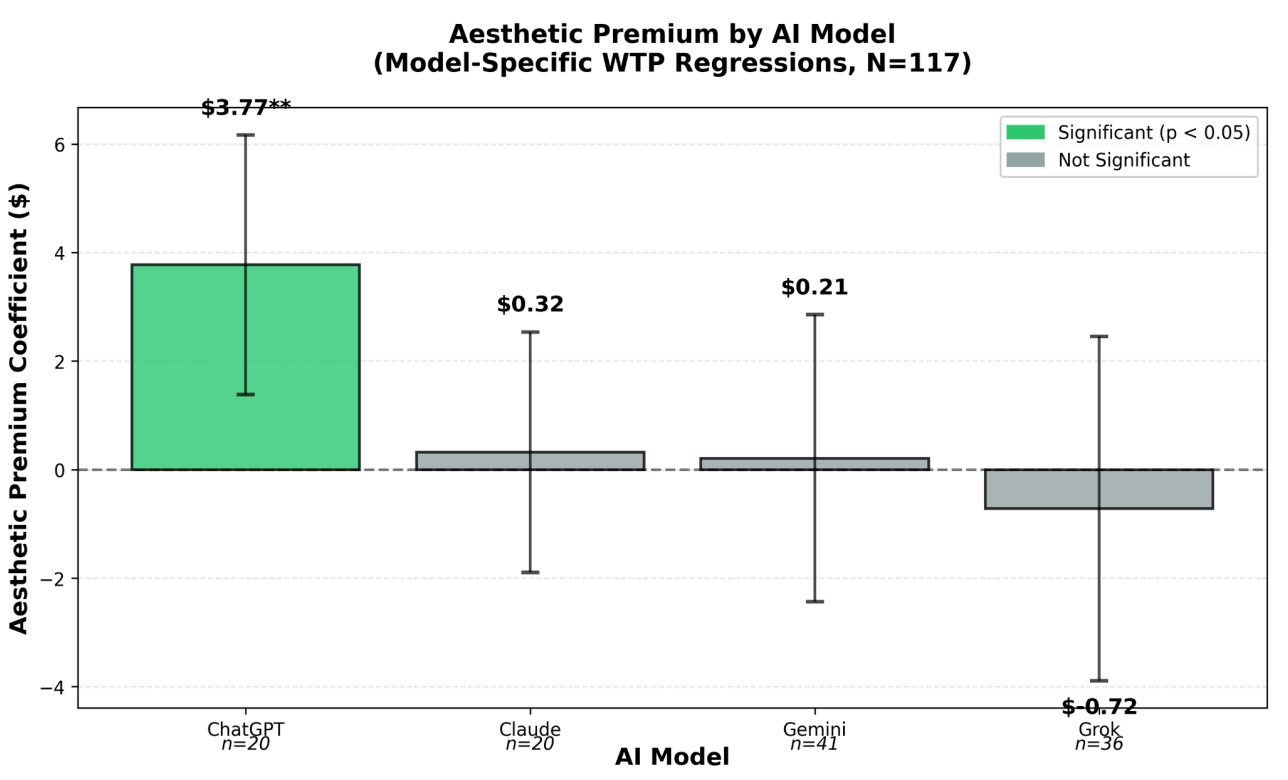

Figure 6: Model-Specific WTP Regressions

## 4.4 Brand-Specific Profiles and Valuation

The brand-specific performance of each individual LLM reveals that the market is deeply polarized. However, the LLM market is not homogeneous amongst providers as there is a rivalry between different "personalities" coded in the fine-tuning of each model. The de-anonymizing of the models (blinded for the rating phase, tracked for analysis) shows different profiles of choice and valuation that point to potential "Superstar" effects.

### 4.4.1 Model Profiles & Selection Bias

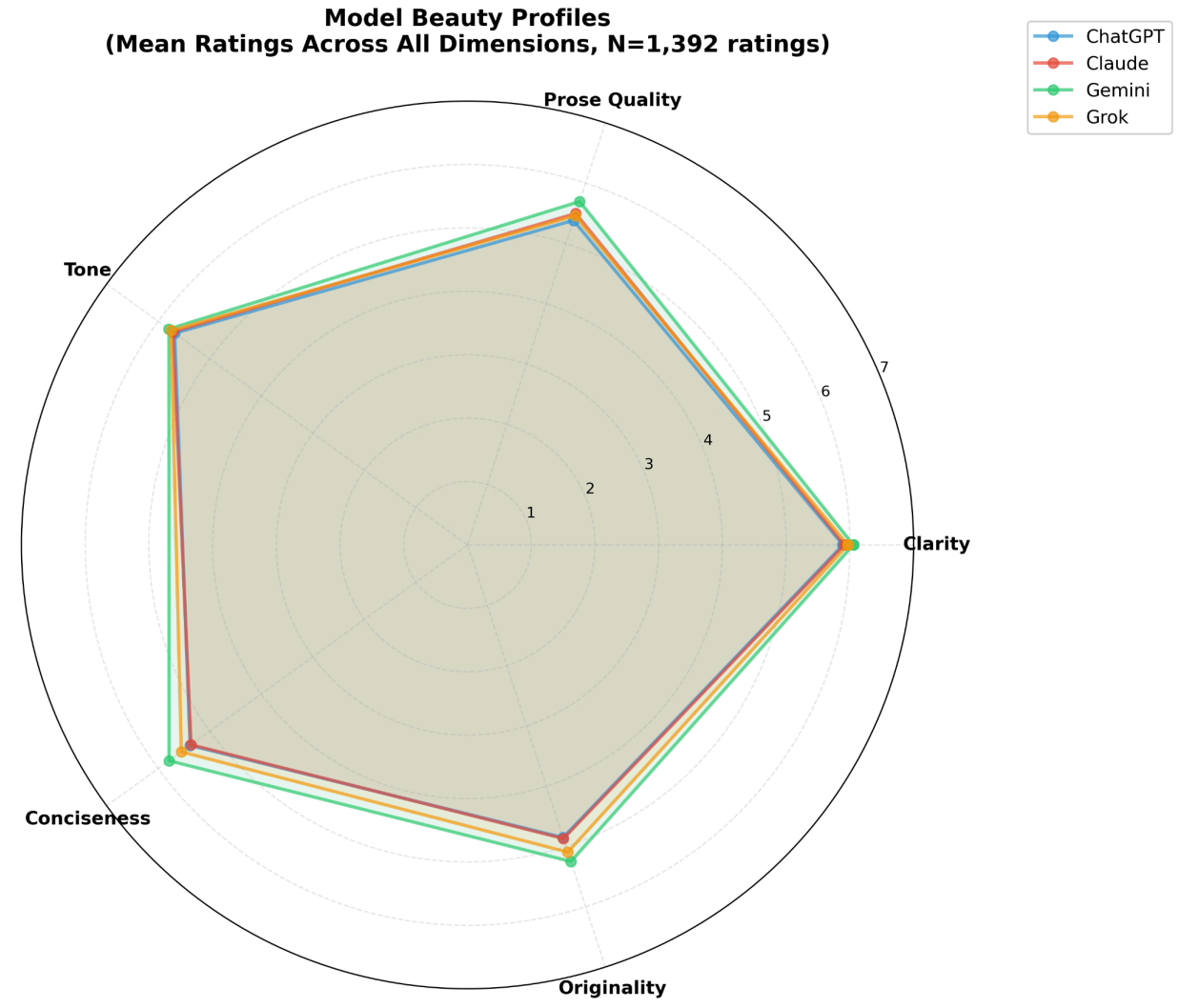

Figure 7: Individual Model Mean Ratings Across all Dimensions

Each LLM exhibits a broadly similar distribution of beauty trait ratings and minimal variation across models, according to Figure 7. This convergence suggests that model outputs are not sharply differentiated in perceived beauty, reinforcing the latent factor structure identified in the PCA (Section 5) and supporting the null effect of beauty framing on willingness to pay discussed earlier in this section.

While all models operate on similar transformer architectures, their Reinforcement Learning from Human Feedback (RLHF) regimes have created a few key measurable stylistic divergences, which Figure 7 reflects. Gemini appears to act as a "utilitarian" character with consistently high scores in functional dimensions like Clarity, but rarely leading in Prose Quality. ChatGPT displays a balanced profile with higher variance in Originality and Tone, suggesting a fine-tuning regime optimized for "human-like" conversational versatility. Grok and Claude show profiles that are often indistinguishable from the mean in the aggregate, though Grok leans towards Originality at the cost of tonal consistency.

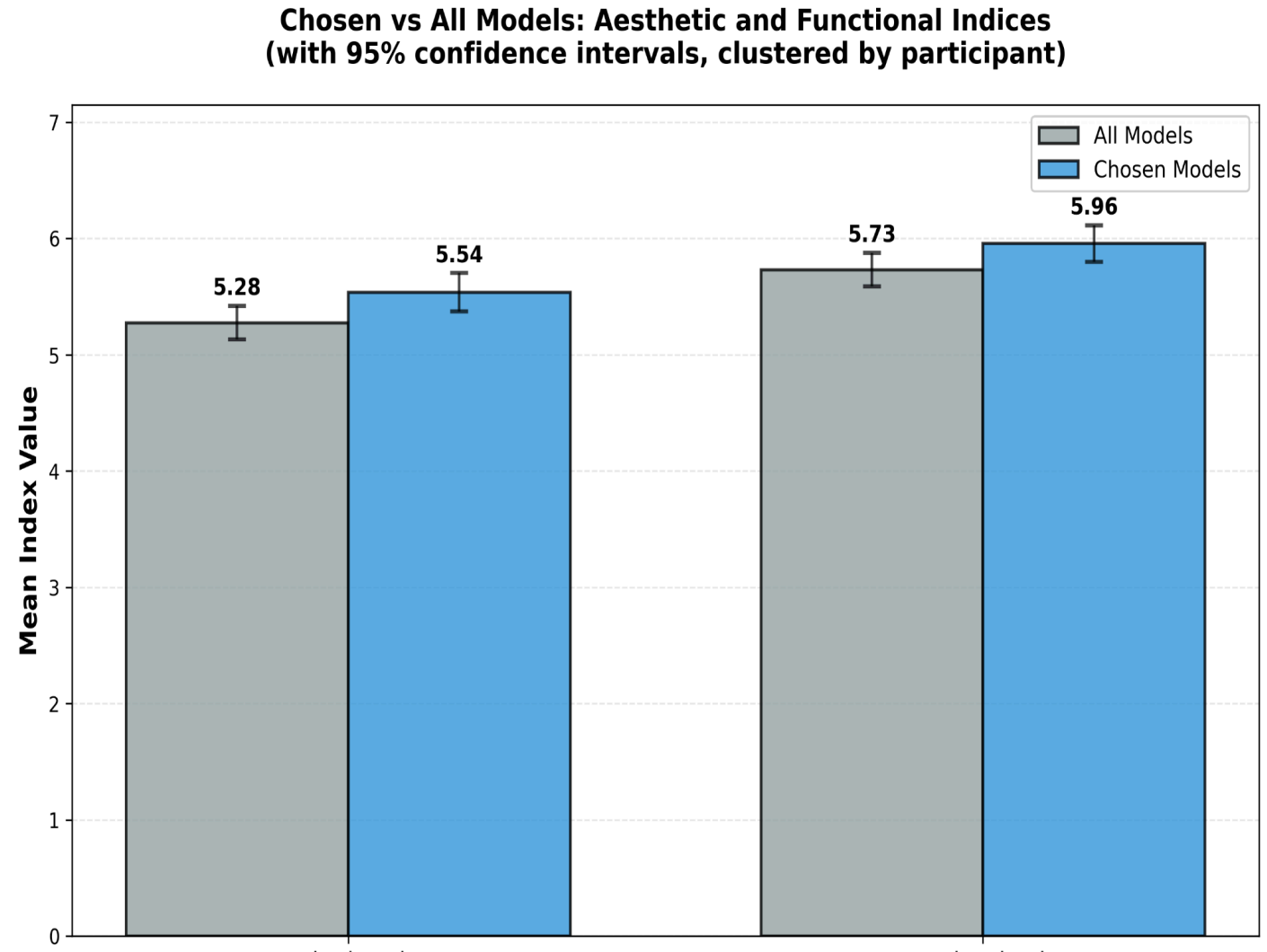

Figure 8: Aesthetic and Functional Indices

The selection mechanism is clarified in Figure 8. When comparing the ratings of models chosen by users against the average ratings of all models, we see a clear selection bias. Users consistently choose models that rate higher on the Aesthetic Index than the pool average. However, the magnitude of this gap varies by brand, suggesting that for some brands, users are more forgiving of lower aesthetic scores, while for others, they demand near perfection.

### 4.4.2 The Divergence of Volume and Value

A particularly notable finding in our brand-specific analysis is the disconnect between Choice Share (popularity) and Economic Valuation (willingness to pay). As detailed in Figure 9, Gemini dominates the choice share, being selected by participants 35.0% of the time. This aligns with its profile as a reliable, functional tool. However, despite this popularity, the aesthetic premium for Gemini is statistically null ($p = 0.861$). Users choose it often, but they do not pay extra for its aesthetic qualities. It functions as the "default" option.

ChatGPT holds a much smaller choice share (17.1%), tied with Claude for last place in terms of selection volume. However, when users do choose ChatGPT, their valuation behavior changes dramatically. The aesthetic premium coefficient for ChatGPT is large and statistically significant ($\beta$ = \$3.774, p = 0.024). A one-unit increase in the aesthetic quality of a ChatGPT output drives a nearly \$4.00 increase in willingness to pay. They are the only model in our study that exhibited a premium of this nature. This suggests that ChatGPT has established a "Brand Aesthetic" where users perceive its style as a luxury good.

Grok and Claude occupy the middle ground. Grok is chosen frequently (30.8%), possibly due to novelty, but its aesthetic coefficient is negative ($\beta$ = -0.717), implying that users choosing Grok may be looking for unobserved stylistic preferences that our rating dimensions do not capture and thus penalize conventional aesthetic standards.

| Model | Choice share (%) | Mean WTP ($) | Aesthetic premium ($) | p-value |
|---|---|---|---|---|
| ChatGPT | 17.1 | 5.50 | 3.774* | 0.024 |
| Gemini | 35.0 | 6.50 | 0.209 | 0.861 |
| Grok | 30.8 | 8.10 | −0.717 | 0.583 |
| Claude | 17.1 | 5.90 | $0.323 | 0.790 |

Figure 9: Model-Specific Economic Indicators

## 5 Defining Beauty in LLM Outputs

The absence of an economic premium for beauty raises a fundamental question about the nature of the attribute itself. If users can rate beauty but won't pay for it, are they distinguishing it from function at all? To answer this, we turn to the psychometric structure of our data.

### 5.1 Unidimensionality in Factor Analysis

A key question raised by our findings is whether users can distinguish beauty from function. If the Lancaster (1966) model holds, users should perceive "Prose Quality" and "Clarity" as distinct characteristics.

We ran principal components analysis on the five standardized textual dimensions (clarity, prose quality, tone, conciseness, and originality), using an unrotated solution. The results visualized in Figure 10 reveal a striking unidimensionality. The Kaiser criterion identifies only one meaningful component (eigenvalue > 3), which explains 63.2% of the total variance. Crucially, a two-factor solution with varimax rotation produced a similar pattern: "Prose Quality" (loading 0.476) and "Clarity" (loading 0.449) load almost identically on this single factor. The other dimensions of Tone, Conciseness, and Originality also load heavily on this first component.

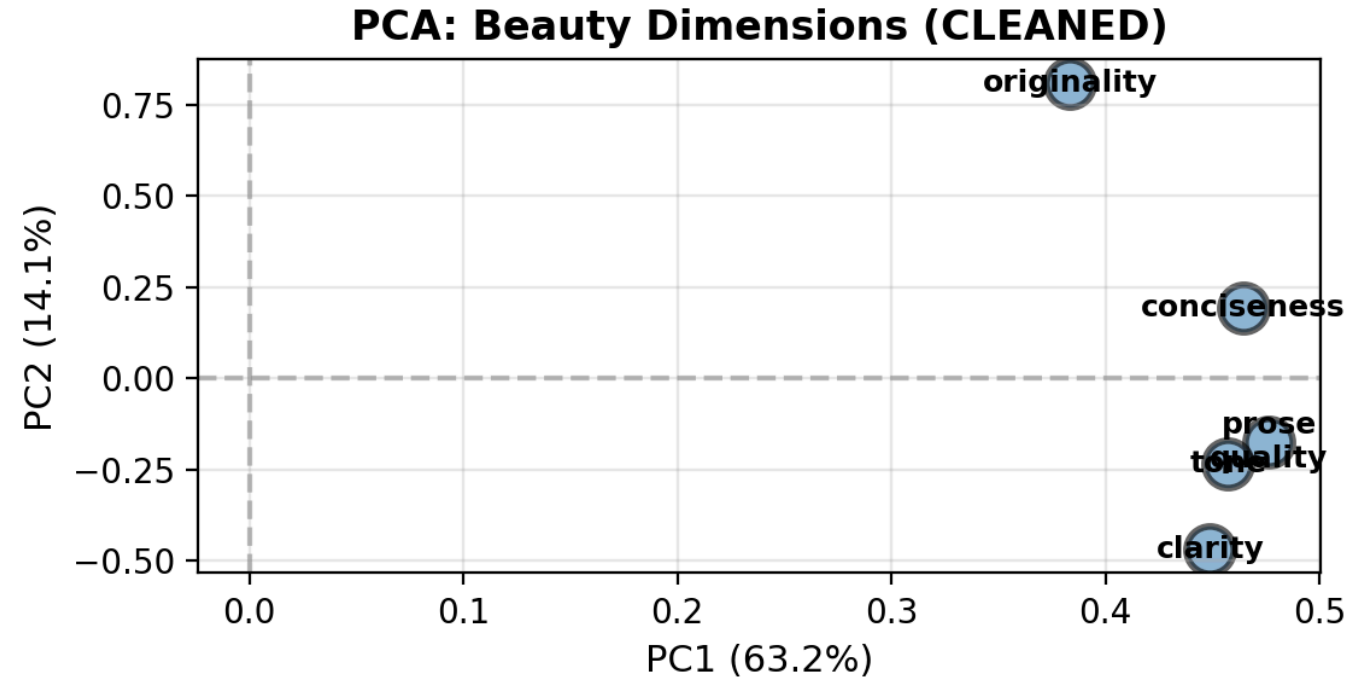

Figure 10: Principal Components Analysis of Beauty Dimensions

In the context of AI text, users do not separate the aesthetic from the functional. A clear text is perceived as well-written; a well-written text is perceived as clear. There is no "beautiful but unclear" quadrant in the user's mental map. Our data reveals that the effective bundle contains only one item: "Goodness." This collapse of dimensions explains why the aesthetic premium is null; there is no separate "aesthetic" variable for the user to value independently of function.

### 5.2 The Definition Reversal

To further explore the mechanism of choice, we examined the drivers of model selection under different experimental frames. This analysis uncovered a counterintuitive "Definition Reversal."

We hypothesized that in the "Beauty" frame, users would prioritize "Prose Quality" and "Originality." Instead, logistic regression, which is reflected in Figure 11, reveals that the strongest predictor of choice in the Beauty frame is Conciseness (β = 0.331). Users instructed to find the most beautiful text gravitated toward the shortest, most efficient outputs.

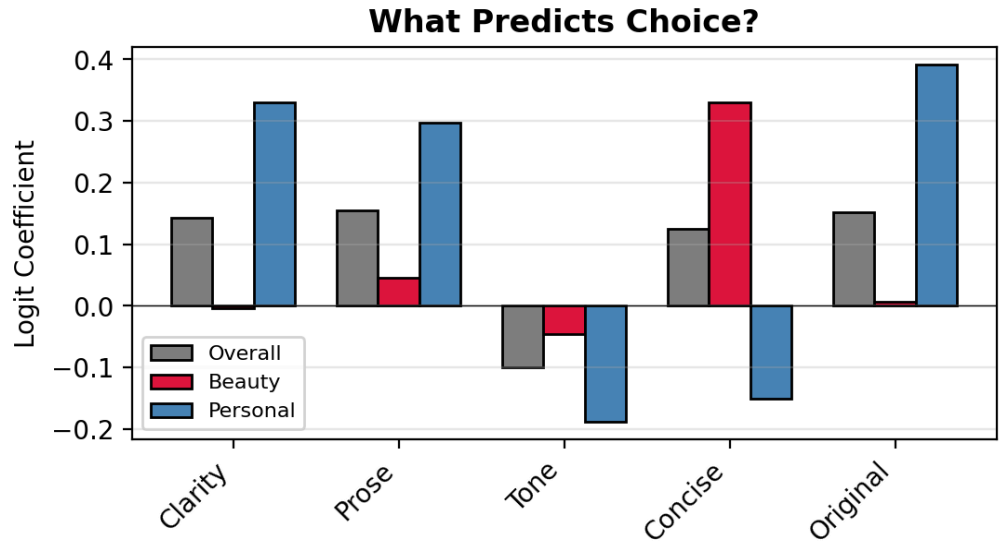

Figure 11: Logistic Regression of Beauty Determinants

Conversely, in the "Personal-Use" frame, Figure 11 indicates the strongest predictor was Originality (β = 0.391, p = 0.084). When choosing for themselves, users valued novelty and distinctiveness. This is a reversal of the expected order. We typically associate "personal use" with utility and "beauty" with creativity. Here, we see the opposite: the abstract search for beauty leads to efficiency, while the personal search for utility leads to creativity. This mirrors the analysis presented in "Conversational Change" by Brynjolfsson et al. (2023)[16], which found that the adoption of AI-driven communication assistants and tools appeared to homogenize the style, length, and content of digital exchanges, leading to a reduction in the idiosyncratic variations traditionally observed across different departments or roles. The observed convergence of communication styles implies a potential trade-off between increased efficiency and the loss of distinctive, human-driven communication styles, but our participant data suggest that one trend transcendent of use-case context is that “beauty” in LLM-generated output is writing that sounds like a concise expert. While the aesthetic ideal of AI converges toward efficient expertise, users seem primed to prioritize the dimensions of novelty and idiosyncratic human-like connection in the personal sphere, preserving creativity as a distinct domain of utility.

## 6 Secondary Outcomes: Choice vs. Price

A divergence emerges between choice and valuation. While the aesthetic ratings do not predict the amount users bid (the intensive margin), they do weakly predict which model is chosen. However, even this relationship is tenuous. The logistic regressions show that the overall predictive power of ratings on choice is low. "Prose Quality" is the strongest overall predictor, yet it fails to reach statistical significance in the aggregate model.

The statistical insignificance of a correlatory relationship between ratings and WTP suggests a "Preference-Valuation Gap." Users may prefer a model because of its style (hence they choose it), but this preference is weak and does not translate into a strong economic commitment. This supports the hypothesis that while exponential progress has democratized high-quality text, it has also demonetized it. High quality is now the baseline expectation, not a value-add. The lack of marginal economic value for differentiated text may have significant implications for the business models of AI companies. If users view high-quality text as a commodity with a reservation price of zero, monetization strategies based on "premium" text quality may fail.

## Conclusion

Across specifications, we find no statistically significant relationship between perceived aesthetic quality and willingness to pay for AI-generated text. Participants consistently distinguish between outputs along stylistic dimensions, but these distinctions do not translate into higher monetary valuation.

For the frontier labs such as OpenAI, Anthropic, Google, and xAI, our results present a strategic dilemma. The industry is currently investing billions in Reinforcement Learning from Human Feedback (RLHF) to fine-tune the "personality," "voice," and "style" of their models. Our findings suggest that these investments successfully smooth the rough edges of AI interaction, raising the baseline quality of outputs across models. However, users expect competence and courtesy as a baseline expectation;

they do not pay extra for eloquence.

This result has direct implications for hedonic models of valuation. In contrast to settings where attributes are separable and command distinct marginal prices, our findings suggest that users do not treat aesthetic and functional qualities of text as independent dimensions. Instead, both collapse into a single perceived measure of quality, limiting the ability of markets to price aesthetic variation independently.

Consequently, the current arms race to optimize for "good writing" may face a commodity trap. To capture economic value, labs may need to pivot from general-purpose aesthetics to highly specialized, context-aware personas that offer distinct functional utility. Another seemingly popular answer is that differentiation will come from agentic capabilities rather than how beautifully it speaks. Ultimately, the economic value of aesthetics in AI systems is contingent on how users conceptualize the role of the technology. As long as language models are treated primarily as tools for task completion, stylistic improvements are likely to be absorbed into baseline expectations rather than monetized as distinct product features

## Acknowledgements

We thank Dr. Eric Maskin, Dr. Amartya Sen, Dr. Abhijit Banerjee, and Dr. Barry Mazur for their help on experiment design and identifying theoretical implications. We thank Prolific and our study participants for providing their time and insight. We thank Mercor for their financial contributions and the advice provided by Dr. Bertie Vidgen, making this research possible.